# Facial Age Estimation: A Research Roadmap for Technological and Legal Development and Deployment

# A White Paper

*Richard Guest*
University of Southampton, UK
r.m.guest@soton.ac.uk

*Eva Lievens*
Ghent University, Belgium

*Martin Sas*
KU Leuven, Belgium

*Elena Botoeva*
University of Kent, UK

*Temitope Adeyemo*
University of Southampton, UK and Ingenium Biometric Laboratories, UK

*Valerie Verdoodt*
Ghent University, Belgium

*Elora Fernandes*
KU Leuven, Belgium

*Chris Allgrove*
Ingenium Biometric Laboratories, UK

*June 2025*

*The authors gratefully acknowledge the support of the 3i University Network in the production of this paper.*

## 1. Introduction

Automated assessment of facial images can provide a technology-based solution for verifying or attributing a person's age. The operation of these systems are broadly in either an **estimation** mode, where a system attributes an age (or a range of ages) to a person without prior knowledge of the subject solely based on facial characteristics or in a **verification** mode where an automated assessment is made against a claimed age [1].

Systems may be deployed to support or automate the assessment of age in cases where a subject is attempting to access age-restricted services or material. The application for this technology may range from remote assessment for the purchase of goods, access to age-restricted social media and on-line content, to the assessment of human subjects for forensic operation or age estimation in undocumented refugee situations. In other situations, age estimation might be used to tailor certain services to the age of the user, such as healthcare, financial services or advertising [2].

Whilst it is important to enhance the technological accuracy of systems, it is vital that deployment is considered across disciplines, encompassing legal, ethical, sociological, and technological spheres and, importantly, the interface between them.

Focusing on age estimation systems, in this white paper we aim to review:

- The current technological issues in age estimation system deployment.
- The legal and regulatory framework within which the technologies operate.
- Future research into the development of, and regulation for robust, fair, and ethical facial age estimation.

## 2. Current deployment of systems and the future applications

The demand for age estimation technologies is spurred on by the increase in remote services that demand age assurance for access to restricted product purchases and other material, particularly in the post-COVID on-line economy, ensuring that the age of the purchaser or user is as claimed. Likewise, age estimation technologies can also be deployed in a variety of applications where age-related services are to be offered and deployed [3].

Table 1 shows example application areas for age estimation systems.

| Application | Description |
|---|---|
| **Security, Access Control, and Law Enforcement** | |
| On-line Product Purchasing/Access | Ensures compliance with legal age requirements in various scenarios. |
| Border Security | Estimate traveller age (particularly with respect to undocumented cases). |
| Law Enforcement and Forensics | Assists identifying missing persons or suspects, particularly in age-sensitive cases. |
| Crisis Management and Humanitarian Aid | Prioritises aid for vulnerable populations during disasters. |
| **Healthcare, Wellness, and Telemedicine** | |
| Geriatric Care and Telemedicine | Monitors aging and enhances remote diagnosis by tailoring medical advice. |
| Fitness Applications | Personalizes workout and diet plans according to biological age. |
| Public Health and Demographics | Aids in population studies and public health planning with accurate age data. |
| **Education and Learning Platforms** | |
| Adaptive Learning Systems | Customises educational content to match learner's age/development stage. |
| Parental Controls | Restricts access to content based on age ensuring safe learning environment. |
| **Insurance, Financial Services, and Legal Systems** | |
| Risk Assessment | Validates age to prevent fraud and accurately classify risk. |
| Customer Engagement | Provides age-specific financial advice and services. |
| **Social media, Gaming, and Entertainment** | |
| Filters and Effects | Enhances user interaction with age-related filters on social media. |
| Age-Appropriate Content | Ensures video games and VR content is suitable for the user's age. |
| Parental Controls in Gaming | Automatically restricts access to certain features based on estimated age. |

*Table 1: Example Age Estimation Application Areas*

## 3. Implementation of facial age estimation systems

### 3.1 Assessment systems

Automated age estimation systems operate by analysis of captured facial images. Many developments in the field have a basis in facial biometric systems – their output is, however, fundamentally different. Whereas biometric systems aim to verify or identify a subject from the facial image sample, age estimation systems attempt to assign an age (or an age band) to a human subject under assessment. The commonality between the operation of the two systems is the extraction of facial features from the image to form a classification. Conventionally, features are typically measurable facial characteristics which are used to train (and subsequently be assessed by) a classifier which utilise multi-dimensional decision boundaries based on input features, which, given the numerical availability of individual features, are explainable in their decision-making process.

Artificial intelligence and, in particular, deep-learning neural systems have revolutionised performance across many domains including age estimation systems. Deep-learning neural systems use multiple layers of numerical processing nodes inspired by the human brain to learn from data. These networks can process information and make predictions. In a supervised training mode, these implementations learn characteristics of presented training data ascribed to a particular class or classes. The internal topologies of these neural systems are typically large and are often presented as pre-trained systems where the operation of the classifier is opaque or a "black box" closed system. The performance of such systems is reliant on the quantity and quality of the data, and particularly demographic diversity represented. Failure to ensure appropriate representation within the training set may lead to a performance bias within a particular population subgroup [4].

The main categories of age estimation technologies are:

- *Traditional Feature-Based Methods* - early implementations of facial age estimation employed handcrafted features such as wrinkle density, skin texture, and facial geometry. Techniques like Local Binary Patterns (LBP), Active Appearance Models (AAM), and Gabor filters were widely used in conjunction with standard machine learning algorithms (e.g. Support Vector Machines, Random Forests). These models benefit from greater transparency but generally lack the performance and adaptability of modern systems, especially in unconstrained environments [5].
- *Deep Learning-Based Approaches* - the emergence of deep convolutional neural networks (CNNs) has significantly improved accuracy in facial analysis tasks, including age estimation. CNNs such as VGGNet, ResNet, and EfficientNet are trained end-to-end to learn age-relevant features directly from image data. More recent developments have adopted transformer-based architectures (e.g. Vision Transformers) and multi-task learning, where age prediction is optimised alongside auxiliary tasks such as gender classification or facial expression recognition [6, 7].
- *Hybrid and Ensemble Models* - Combining multiple models either using different architectures or integrating feature types has proven effective in improving robustness. These ensemble methods reduce variance and can adapt to diverse input conditions, such as varying lighting, resolution, and pose. For example, systems may average predictions from models trained on aligned and non-aligned facial inputs or aggregate predictions across multiple facial regions [8].

### *3.2 System testing and evaluation*

Accurate testing of age estimation systems is important to ensure that stakeholders and, most importantly, the public have confidence in their operation. Testing of age estimation systems lags behind biometric verification systems testing protocols and standardisation. PAS 1296:2018 [9] from the British Standard Institute provides some guidance for testing, whilst ISO/IEC have a standard (ISO/IEC 27566-1) in draft development as of 2025 [10] which will establish "a framework and core characteristics for age assurance systems deployed for the purpose of enabling age-related eligibility decisions" including system testing methods.

Testing involves presenting images to a system containing faces of subjects with a known ground-truth age. Testing of age estimation systems relies on using a demographically representative dataset of images, including age, gender and skin tone. Given the age boundaries under consideration, this includes images of children as well as adult subjects. Images should also be diverse in terms of pose, location, head coverings, make up, glasses, etc. The US National Institute of Standards and Technology (NIST) conducted the first independent large-scale trial of age estimation systems in their Face Analysis Technology Evaluation (FATE-AEV) programme, with the latest results released in 2025 [1]. They indicated that the introduction of neural systems accounted for a significant performance increase since the first generic facial tests a decade previously.

For age estimation systems, the issue of training (and testing) data is problematic. Given that judgement decision boundaries for age-restricted access (for example, at ages 12-13, 15-16, 17-18, 20-21 and 24-25+) are compressed at the lower end of the population distribution, and rely on utilising facial images of child subjects, access to large datasets across these decision boundaries is highly restricted. Despite high accuracy of age estimation algorithms under ideal conditions, real-world deployment introduces a host of challenges:

- *Image Quality Variability:* Uncontrolled lighting, background noise, image resolution, and camera angles can reduce model accuracy. Mobile devices, webcams, and low-end sensors often produce suboptimal input.
- *Appearance Alteration:* The ability for within-person short-term appearance modification through cosmetics, facial hair, clothing/accessories (glasses, hats), and intentional disguise can mislead both traditional classifiers and neural models. Temporary physical changes (e.g., fatigue, weight gain, health issues) further complicate predictions.
- *Adversarial Vulnerabilities:* Deep neural networks are susceptible to adversarial attacks, where imperceptible changes to an input image can lead to drastically incorrect age estimates [12]. Attackers may also use deepfakes or morphed images to bypass systems.
- *Demographic Bias and Fairness:* Disparities in performance have been documented across gender, ethnicity, and age groups. Underrepresentation in training data leads to systematic errors, particularly for minority populations and individuals at age extremes [13].
- *Remote Capture:* When facial sample donation occurs in an unsupervised remote scenario the process is open to attack through presentation spoofing using conventional photographic instruments. It is also possible for siblings to form a replacement attack by donating a facial sample against a reference image with strong similarities.

### *3.3 Performance metrics*

The two main processes for assessment include calculating a) the algorithmic performance by assessing the difference between estimated and actual age for individual images and b) assessing the operational accuracy of a system when applying an age threshold (typically setting a threshold at a legal age boundary).

<u>*Age Estimation Model Performance*</u> assesses how accurately the algorithm/model predicts age values, independent of any thresholding or decision-making process, focusing purely on the model's age prediction performance.

<u>A*ge Estimation System Testing*</u> evaluates the overall system's effectiveness in real-world use cases, focusing on age-related eligibility decisions (e.g. "Is this person over 18?"), validating the reliability and trustworthiness of the complete system, including data handling, context, and operational use.  Common performance metrics include:

- *MAE (Mean Absolute Error)*: Average of absolute differences between predicted and actual ages.
- *RMSE (Root Mean Squared Error):* Square root of average squared differences which penalises larger errors.
- *ME (Mean Error):* Average signed difference thereby useful for detecting bias - overestimation or underestimation.
- *MedAE (Median Absolute Error):* Median of absolute errors; less sensitive to outliers than MAE.
- *SDE (Standard Deviation of Errors)*: Indicates variability in age predictions, assessing model consistency.
- *Threshold Accuracy*: Percentage of age predictions correctly placed with respect to a decision boundary. Precision, Recall and F1 can also be applied to show performance in detecting 'true' results (where 'true' is represents samples/predictions to one side of a decision boundary).
- *PTA (Prediction Tolerance Accuracy):* Percentage of predictions within a defined error threshold (e.g., ±2 years).
- *APA (Accuracy Per Age)*: Proportion of accurate predictions for each specific age (e.g. how often 17-year-olds are correctly identified as 17).
- *PET (Percentage Error Threshold):* Maximum allowable error that still captures a defined percentage of the dataset (e.g., 90% of predictions fall within ±3 years).

*3.4 Explainability*

Alongside the technological development of accurate age estimation systems is the requirement for systems to be explainable in their decision-making process – what were the characteristics that resulted in a particular age being returned by the system? As noted, many state-of-the-art solutions utilise deep learning AI architectures, which require the use of post-hoc explainable AI (xAI) techniques for their analysis. Feature attribution methods identify the input features that contribute most to the overall classification - LIME and SHAP are widely used as xAI methods that produce so called "surrogate" interpretable models that locally approximate the original model and hence can identify where decisions are made by the hidden "black box" architecture. As an alternative, a saliency map provides a visual representation that highlights the parts of an input (for example, a facial image) that most influenced a machine learning model's prediction. However, for image data, these kinds of explanations may not be as informative as they may not represent describable concepts understandable to humans. An alternative solution is to utilise AI architectures that are explainable by design. Yet, explainability-by design may create vulnerabilities to attacks because their functioning is transparent and therefore easier to circumvent [14].

## 4. The legal and regulatory framework for facial age estimation

From a legal perspective, age estimation is relevant as a mechanism that is used to implement legal restrictions on access to certain services or goods to individuals of certain ages (usually children of varying ages), on the one hand, and as a technology that needs to comply with existing legal requirements imposed by different instruments (e.g. data protection regulation as age estimation technology processes personal data), on the other hand. In this section both aspects are described. Scenarios in which age estimation is used to tailor services to the age of individuals (such as healthcare, financial information or advertising) are not discussed below, although, in the EU, such applications will also need to comply with legal instruments, such as the EU General Data Protection Regulation (GDPR).

*4.1 Age assurance and (facial) age estimation in law and policy*

Age assurance is increasingly encouraged or mandated by law,[1] either explicitly or implicitly, and often as a tool to achieve better protection of children in the online environment. At the EU level, certain legislative instruments suggest age assurance as a protective measure – such as Article 35(1)(j) of the EU Digital Services Act (DSA) and Articles 6a and 28b of the EU Audiovisual Media Services Directive (AVMSD) without explicitly requiring its use. Implicit legal bases for age assurance can also be found, for instance, in Article 8 General Data Protection Regulation (GDPR). These legal instruments often refer to "age verification" or "age assurance" but never point to specific methods. Hence, facial estimation is nowhere explicitly mentioned. In recent guidance documents, such as the European Data Protection Board (EDPB)'s statement on Age Assurance [15] and the European Commission's draft guidelines on measures to ensure a high level of privacy, safety and security for minors online pursuant to Article 28(4) DSA [11] there are references to age estimation as an age assurance method, but not to facial age estimation in particular. By contrast, national legislation may impose more specific obligations, for example by mandating age limits for performing legal acts or acquiring certain goods or services [16].

---

[1] Shaffique & van der Hof explain that age assurance is legally relevant in three ways "(1) when a minimum age is prescribed by law for buying products or using services that may harm children or for performing legal acts, both of which require age assurance for legal compliance, (2) when there is a duty of care to protect children, which may require age assurance to be employed, and (3) when there is a contractual obligation to provide the products or services only to users of a certain minimum or maximum age" [16].

### *4.2 Risks associated with facial age estimation*

Recent research [17, 18] has identified the legal challenges associated with age assurance in detail. Specifically with regards to age estimation, Sas & Mühlberg conclude that there are risks related to privacy intrusion, inaccuracy and discrimination (due to biases in the training data set) and classification errors leading to individuals being blocked from access to services to which they are actually entitled to. Furthermore, depending on where the analysis is run (locally on the user's device or remotely), there might be security risks (e.g. data leaks). Such risks may impact a range of fundamental rights, such as the right to privacy and data protection and the right to non-discrimination. While this is true for individuals of all ages, particular concerns relate to the impact on the rights of children.

### *4.3 Legal requirements and compliance for facial age estimation*

Children's rights
While age assurance is acknowledged to be an important tool in achieving a higher level of protecting children, at the same time, the risks reiterated above in relation to facial estimation do have the potential to affect rights that are attributed to children in the UN Convention on the Rights of the Child (UNCRC), such as the right to privacy and non-discrimination. During consultations for the recent evaluation of the Better Internet for Kids+ Strategy, many children and young people (aged 6-18) voiced concerns about privacy and data protection in relation to the age assurance methods used on the websites, platforms, and apps they regularly access [19].

When children are the affected individuals, scholars point to the importance of conducting a Child Rights Impact Assessment (CRIA) [17, 18] before deploying age assurance mechanisms, for instance, to understand which methods are proportionate and do not negatively impact on the full range of children's rights.

General Data Protection Regulation

As facial estimation techniques process personal data, data controllers – those who decide on the purpose and means for the processing – will need to comply with the obligations laid down in the GDPR. This includes compliance with the data protection principles in Article 5 GDPR, which encompass, amongst other areas, lawfulness, data minimisation, accuracy, and security. Article 25 GDPR which requires the observance of data protection by design and default[2] is relevant as well. Examples of measures that could be taken to ensure compliance with the GDPR include absence of identification, prompt deletion of facial images, on-device processing and encryption.

In addition, as age assurance often targets children, specific protection for the processing of their personal data must be ensured (recital 38 GDPR). According to the EDPB, service providers must evaluate the necessity and proportionality of age estimation mechanisms concerning children's rights. This assessment can be carried out through Children's Rights Impact Assessments (CRIAs; see above), which may be integrated into the mandatory Data Protection Impact Assessment (DPIA) (Article 35, GDPR) in this context [16].

An important question which arises is whether facial estimation entails the processing of a special category of data or 'sensitive' data, according to article 9 of the GDPR. Biometric data only falls within the scope of the prohibition for processing sensitive data when it is used for the purpose of *uniquely identifying* a natural person, which is not the case for facial age estimation techniques. Yet, at the same time, as Sas & Mühlberg argue, facial images may disclose information which does fall within the special category of personal data under article 9 GDPR (e.g., racial or ethnicity origin, or information about health conditions or disabilities affecting facial appearance), and, hence, they too deserve heightened privacy protections [18].

Artificial Intelligence Act

As facial estimation techniques may be classified as AI systems according to the 2024 EU Artificial Intelligence Act (AIA), further obligations may arise for providers and deployers of such systems. The most relevant question is whether facial age estimation systems could be qualified as a high-risk system indicated in Annex III to the AIA. This could be the case insofar as facial age estimation would be considered an AI system "intended to be used for biometric categorisation, according to sensitive or protected attributes or characteristics based on the inference of those attributes or characteristics" (Annex III, 1(b), AIA). While it could be argued that facial age estimation does categorise individuals according to age, whether it does so according to sensitive attributes or characteristics based on the inference of those attributes is perhaps less clear. Other doubts may arise when reading recital 54 AIA which states that "[b]iometric systems which are intended to be used solely for the purpose of enabling cybersecurity and personal data protection measures should not be considered to be high-risk AI systems". In any case, if facial age estimation would be considered a high-risk AI system a host of obligations (including risk management)[3] need to be complied with. If such a system is deployed by bodies governed by public law, or if private entities are providing public services, a fundamental rights impact assessment needs to be carried out as well (Article 27 AIA).

---

[2] See [19]. In this decision the European Data Protection Board the EDPB expresses serious doubts about the effectiveness of the self-reported age mechanism for high-risk processing and as to whether TTL provided sufficient evidence that the measures in place reduced the likelihood of children under the age of 13 gaining access to and being active on the TikTok platform. Notwithstanding these serious doubts, the EDPB concludes that it does not have sufficient information to conclusively assess compliance with Article 25(1) and, hence, cannot find an infringement. See also [20].

### *4.4 Remedies for individuals*

In cases where the above-mentioned rules are not respected, individuals have certain remedies at their disposal. This includes submitting a complaint with a Data Protection Authority (Article 77 GDPR) or a Market Surveillance Authority (Article 85 AIA) or obtaining from the deployer clear and meaningful explanations of the role of the AI system in the decision-making procedure and the main elements of the decision taken (Article 86 AIA). However, very often data subjects or individuals who are affected by AI systems will not be aware of potential infringements due to the opaque nature or functioning of these systems. Furthermore, additional difficulties may arise when the affected individuals are children, as they are often even less aware and will need assistance when exercising their rights.

### *4.5 Recent developments regarding standards and guidance*

Given that legal instruments remain silent on the specific methods that should be used for age assurance, despite the growing prominence of the issue in public and policy debates and the significant concerns about the risks associated with these techniques, there is an urgent need for more guidance. At EU level, a Task Force on Age Verification under the Digital Services Act was set up in January 2024 [22]. The European Commission's draft guidelines on measures to ensure a high level of privacy, safety and security for minors online pursuant to Article 28(4) DSA [11], made available for consultation in May 2025, refer to age estimation an appropriate method to give access to platforms to minors when their terms and conditions require a minimum age lower than 18 to access the service, or when they offer 'medium risk services'. A prior proportionality assessment is always required and features such as accuracy, reliability, robustness, non-intrusiveness, and non-discrimination must be taken into account.

---

[3] Specifically, Article 9.9 AIA requires that when implementing a risk management system, providers shall give consideration to whether in view of its intended purpose the high-risk AI system is likely to have an adverse impact on persons under the age of 18.

At the national level, regulatory bodies have also issued guidance: in the UK, Ofcom has published detailed guidance on highly effective age assurance [23]; Arcom in France has developed technical guidelines for age verification to protect minors from online pornography [24]; and Spain's AEPD has issued both a Decalogue of principles and a technical proof of concept for age verification systems [25]. The UK Government has also established the Age Estimation Science Advisory Committee to advise on "the use of existing and emerging scientific approaches" which alongside the Biometrics and Forensics Ethics Committee is advising on implementing age estimation methodologies.

Efforts are also underway to develop technical standards. For instance, the IEEE Online Age Verification Working Group published its 2089.1 Standard for Online Age Verification in May 2024 [26], and ISO is currently developing a draft standard on age assurance (ISO/IEC 27566; see above) [10]. The euConsent project produced a proof of concept for an ecosystem aimed at creating interoperability in the online age assurance market (i.e. its AgeAware Specification which includes business and technical requirements) [27].

These various initiatives reflect a growing need to balance the protective aims of age assurance with strong safeguards for the full range of children's rights including privacy, non-discrimination and participation.

## 5   Future research into the development of and regulation for robust, fair, and ethical facial age estimation.

As shown, age estimation systems have significant potential in various fields, including safety and security, healthcare, retail, and entertainment. At the same time, there are concerns regarding the accuracy of the technology, training processes, explainability, risks to fundamental rights and legal compliance.

This section contains a roadmap outlining the key areas for future exploration in developing, deploying, and refining age estimation systems, guiding stakeholders through the key stages of research, technology selection, implementation, and validation. The roadmap emphasises a combination of data-driven methodologies, ethical and legal considerations, and technological innovation to achieve accurate and fair outcomes.

*5.1 Training age estimation systems*
Collecting facial data to train age estimation systems based on AI presents significant legal and ethical challenges. To ensure that these systems produce accurate results for all groups within a population, it is essential to gather representative data, including from children, minorities, and individuals with distinctive facial features or conditions. However, this necessity creates a paradox: while the goal is to reduce the collection of personal data, achieving accurate age estimation requires collecting and storing more data, which increases the risk of misuse and breaches. This data collection can inadvertently capture sensitive information protected under Article 9 of the GDPR, such as ethnicity, health conditions, and religious attributes. The storage of facial data, even if intended for training purposes, can lead to potential identification and misuse, including impersonation, deepfakes, and targeted profiling. These risks underscore the need for robust safeguards and ethical considerations in handling facial data for training purposes.

Questions to be resolved include:
- How can developers reconcile the need for large and high-quality datasets with the GDPR's principles of data minimisation and purpose limitation? Is it possible to achieve both accuracy and privacy, or does it reflect a fundamental tension in current regulatory frameworks?
- What constitutes a legitimate legal basis for collecting facial data for training purposes, especially when data is sourced from public platforms or third parties? Is informed consent feasible or meaningful in these contexts?
- How can transparency be ensured when real faces are scraped from the internet for AI training purposes, and what ethical and legal boundaries should guide such practices?

*5.2 Synthetic data*
Synthetic data is increasingly valuable for training age estimation systems, addressing key challenges such as data scarcity, bias, and privacy concerns. Age estimation models require diverse and high-quality datasets to perform accurately across all age groups, demographics, and conditions. However, obtaining real-world data, particularly for underrepresented age ranges like children or elderly individuals, is often difficult due to ethical and privacy restrictions. Synthetic data offers a scalable solution by generating realistic images using techniques like Generative AI and 3D modelling. It allows for the creation of balanced datasets with equal representation of genders, ethnicities, and age groups, helping mitigate biases that are common in real-world data. This improves fairness and ensures better generalisation across diverse populations. Additionally, synthetic data can simulate edge cases and controlled variations, such as different lighting, occlusions, and facial expressions, enhancing the model's ability to handle real-world scenarios.

However, the quality and realism of synthetic data are critical. If synthetic data fails to mimic real-world variability, the model may not generalise well. Therefore, combining synthetic data with real data during training could be a more effective approach. The 'quality' and intra-subject diversity of synthetic samples that can be generated by evermore powerful Generative AI techniques represents an on-going research question.

From a legal perspective, it could be argued that the use of synthetic data could be beneficial to the right to privacy of children if no real images need to be collected for training purposes. Yet, at the same time, it might still be the case that synthetic data would fall within the scope of data protection legislation, depending on how images are generated and their qualification as anonymous or personal data. Moreover, the Artificial Intelligence Act could also be of relevance, prompting the question regarding the quality of data sets in relation to high-risk systems. From an ethical perspective, the questions arises whether it is acceptable and justifiable to test age estimation services with a potential significant impact on the rights and lives of children with images of children which do not actually exist.

### *5.3 Performance testing of age estimation systems*
Without rigorous testing, age estimation systems might consistently misestimate ages and thus ensuring they produce accurate and reliable results helps maintain trust and usability. Whilst standards are being developed for the assessment of age estimation systems, there remain research issues in terms of protocol and datasets for repeatability.

In parallel with unresolved issues with the use of synthetic data, detailed in Section 5.2, the responsible and ethical use of genuine data requires investigation. Whereas adult sample population is relatively easy to obtain (although required to be ethical in its collection with appropriate consent and usage), databases of child facial images quite rightly are less easy to obtain due to sensitive ethical issues surrounding the data capture. Accurate ground-truth data across multiple ages and other demographics is necessary for the attribution of the fine-grain age bins. There is also the need to assess the availability of longitudinal datasets that will provide evidence for short, medium and long-term ageing processes.

Age estimation systems can also be biased across demographic groups, such as different genders, ethnicities, or skin tones leading to the potential for underestimates or overestimates ages for specific groups. Real-world environments introduce variability, such as lighting conditions, image quality, and diverse user appearances. Robust and transparent testing should be implemented with diversity of subject demographics and image scenarios to ensure systems perform consistently across different settings and scenarios, ensuring accuracy, fairness, ethical deployment and performance, and reliability.

### *5.4 Presentation attack detection*

Presentation Attack Detection (PAD) is crucial for securing age estimation systems against fraudulent attempts to manipulate age verification. Presentation attacks occur when individuals use techniques like printed photos, video replays, 3D masks, or deepfakes, alongside physical appearance changes such as make-up, clothing, hairstyles/facial hair and glasses to appear older or younger and bypass restrictions. Whilst some of these presentation modifications may be made to attack systems, very often they just represent lifestyle choices in terms of appearance presentations. PAD ensures the system can differentiate between bona-fide and spoofed presentations. Sitting alongside conventional performance testing, the accurate assessment of PAD for age estimation, needs to be ensured, in order to enhance trust of the public in the use of this technology.

### *5.5 Explainability*

As noted, explainable systems allow users and stakeholders to understand how the system arrived at a particular age estimation, building trust especially in high-stakes scenarios such as verifying access to age-restricted services. With explainability, deployers can provide clear reasoning for decisions, reducing suspicions of arbitrary or opaque results. Drawing on, and extending, existing xAI work, the community should work towards the use of understandable explanations of the decision-making, specifically of age estimation processes.

This links to broader debates about AI transparency in judicial settings, where judges, lawyers, and affected individuals must be able to scrutinise and challenge AI-based findings, if disputes arise. One important question here is what standards of transparency and explainability are necessary for facial age estimation systems to be admissible and meaningful as evidence in court? Past initiatives, such as the judicial primers project led by the Leverhulme Centre for Forensic Science at University of Dundee, may serve as useful models for translating complex technologies and methods into comprehensible formats for judges, lawyers and laypersons [28]. Yet much remains to be done, including identifying best practices for documenting AI decision-making processes and making them explainable.

### *5.6 Guidance on legal compliance*

Although facial age estimation technologies fall within the scope of different legal instruments, such as the GDPR and the AIA, many of the legislative provisions give rise to uncertainties and questions regarding the concrete implementation in relation to facial age estimation. Whereas guidance on age assurance in general is increasingly provided both at European and national levels, age estimation – let alone facial age estimation – is often only mentioned in passing and the intricacies of ensuring compliance are left unaddressed.

ƒQuestions to be resolved include:
- Which methodologies are appropriate for legally required impact assessments (e.g. DPIAs, fundamental rights impact assessments, child rights impact assessments) in the context of facial age estimation to mitigate potential negative impacts on fundamental rights?
- How should responsibility for compliance and correct and bias-free age estimation be attributed to the various actors in the chain? How can this be operationalised under existing law like the AIA?
- To what extent should facial data be treated as special category data under Article 9 GDPR, particularly when such data may unintentionally reveal sensitive attributes such as ethnicity, health status or religion?
- Should facial age estimation systems be qualified as high-risk under the AIA, triggering enhanced obligations for providers and deployers? If not, what regulatory gaps remain unaddressed?
- Which legal remedies are available to individuals, especially children, who are adversely affected by incorrect facial age estimation decisions? How accessible and effective are existing avenues, such as regulators, courts, or children's rights commissioners, and do they provide timely and child-sensitive redress?

### 5.7 User trust

Trustworthiness of age estimation systems is essential. Public trust is a prerequisite for the deployment and take-up of facial age estimation technologies in daily life. Trust may be affected by (perceived) accuracy and reliability as well as by legal and ethical concerns, regarding risks and the impact on fundamental rights.

Questions to be resolved include
- What are the views of users, especially children, on the use of facial age estimation? What are their views on convenience, intrusiveness and privacy in this context?
- Can trustworthiness be improved through the provision of accurate and reliable information and education?

It is clear that interdisciplinary solutions and research are required to address these issues with input from psychologists, anthropologists, computer scientists, legal scholars among others.